\documentclass[aps, 
	prd,
	superscriptaddress, 
	preprintnumbers, 
	amsfont,
	amssymb,
	amsmath,
	notitlepage,
	longbibliography,
	nofootinbib]{revtex4-1}
\usepackage[colorlinks=true, 
	linkcolor=blue, 
	citecolor=blue, 
	urlcolor=blue, 
	linktocpage=true]{hyperref}
\usepackage[utf8]{inputenc}
\usepackage[english]{babel}
\usepackage{lmodern}
\usepackage[protrusion=true, expansion=true]{microtype}
\usepackage[dvipsnames]{xcolor}
\usepackage{graphicx}
\usepackage{mathtools}

\newcommand{\lp}{\ensuremath{\left(}}
\newcommand{\rp}{\ensuremath{\right)}}
\newcommand{\s}{\nobreak\hspace{.08em plus .04em}}
\newcommand{\dd}{\ensuremath{\mathrm{d}}}
\newcommand{\e}{\ensuremath{\mathrm{e}}}

\newcommand{\pd}{\partial}

\definecolor{remFcol}{RGB}{0,140,220}

\begin{document}

\title{Relativistic collapse of axion stars}
\date{\today} 
\author{Florent Michel}
\email{florent.c.michel@durham.ac.uk}
\affiliation{Centre for Particle Theory, Durham University,
South Road, Durham, DH1 3LE, UK}
\author{Ian G. Moss}
\email{ian.moss@newcastle.ac.uk}
\affiliation{School of Mathematics, Statistics and Physics, Newcastle University, 
Newcastle Upon Tyne, NE1 7RU, UK}

\begin{abstract} 
We study the gravitational collapse of an axion field in null coordinates, assuming spherical symmetry. 
Compared with previous studies, we use a simpler numerical scheme which can run, for relevant parameters, 
in a few minutes or less on a desktop computer. We use it to accurately determine the domains of parameter 
space in which the axion field forms a black hole, an axion star or a relativistic Bosenova.
\end{abstract}

\maketitle

\section{Introduction}

Amongst the possible dark matter candidates, a coherent scalar field
with very low mass is an enticing possibility. The idea originated with the 
QCD axion \cite{Sikivie:2006ni}, but the concept has since been extended to a class of
axion-like particles (ALP's) with ultra-light masses \cite{Arvanitaki:2009fg}. In ALP scenarios, 
the dark matter forms gravitationally bound objects which may form into
galaxy cores \cite{Schive:2014dra}, or for larger masses into axion mini-clusters 
\cite{Hogan:1988mp,Kolb:1993zz,Seidel:1993zk}. 
These objects are often stable only for a particular mass range, leaving the possibility of 
detectable cosmological signatures from the axion bound structures or from 
the remnants of their collapse.

ALP's are characterised by their mass $m$ and decay constant (or symmetry
breaking scale) $f$.
Coherent ALP dark matter scenarios envision the dark matter energy density 
in the form of large-scale coherent axion oscillations of frequency $\sim m$,
with density parameter \cite{Sikivie:2006ni}
\begin{equation}
\Omega_{\rm ALP}\sim 0.1\left({f\over 10^{17}\,{\rm GeV}}\right)^2
\left({m\over 10^{-22}\,{\rm eV}}\right)^{1/2},\label{omega}
\end{equation}
although this is rather dependent on initial conditions. 
Spatial gradients
in the oscillating axion field induce ``quantum'' pressure forces
which are capable of supporting structures on the Kpc scale
for axion masses  around $m\sim 10^{-22}\,{\rm eV}$, or galaxy
Halo scales for $m\sim 10^{-24}\,{\rm eV}$ \cite{Arvanitaki:2009fg}.

We follow the recent trend of referring to stable axion structures as axion stars
(though the term Bose star is also frequently used in this context). So far
three distinct scenarios of gravitational collapse for APL's have been identified
\cite{Levkov:2016rkk,Helfer:2016ljl}: they can settle down quietly to an axion star; 
they can radiate away energy in bursts of relativistic axions or they can 
collapse to a black hole.
The second outcome is a relativistic analogue of the Bosenova phenomena 
in cold-atom physics \cite{Donley:2001}. Like the cold atoms in a Bosenova, the axions have an 
attractive self-interaction force which can overcome the quantum pressure. We 
will use the term Bosenova in this paper to refer to the axion collapse and radiation
phenomenon.

The fate of an axion clump can be represented on phase
diagrams labelled by parameters describing the axion properties and the initial conditions. Recently, Helfer et al. 
\cite{Helfer:2016ljl} have  produced a phase diagram for spherically symmetric 
collapse with axion decay constant $f$ and the initial mass of the 
axion clump, and they have speculated that there 
is a tricritical point joining phase boundaries between the three outcomes. 
The aim of this paper is to provide convincing numerical evidence for the tricritical 
point using a particularly amenable form of the field equations, and to determine the parameter
values accurately at the phase boundaries.

We use the null-coordinate integration schemes introduced into spherically symmetric
gravitational collapse by Goldwirth and Piran \cite{PhysRevD.36.3575}. 
The null techniques are particularly efficient because the coordinate grid flows
inwards with the collapsing matter. For example, the null methods can reproduce
the universal scaling phenomena in massless scalar collapse  \cite{Garfinkle:1994jb},
which otherwise is only possible with less efficient mesh refinement techniques 
\cite{Choptuik:1992jv}.

Throughout this work, we use units in which the reduced Planck constant 
$\hbar$ and velocity of light $c$ are equal to unity. The reduced Planck mass 
$M_p=(8 \s \pi \s G)^{-1/2}$, where $G$ is Newton's constant.

\section{Model and field equations}

We take the generic axion potential $V$, which is typical of the potentials which represent axion dark 
matter~\cite{Kuster:2008, RevModPhys.82.557, PhysRevD.81.123530}:
\begin{equation}
V(\phi) = m^2 \s f^2 \s \lp 1 - \cos \lp \frac{\phi}{f} \rp \rp.
\end{equation}
The parameters $m$ and $f$ are related by (\ref{omega})
 if the cosmological dark matter density is in the form of coherent axion oscillations, but we will generally
take $m$ and $f$ as free parameters. The Lagrangian density  of the axion field is
\begin{equation} \label{eq:Lphi}
\mathcal{L}_\phi = - \frac{g^{\mu \nu}}{2} \s (\pd_\mu \phi) \s (\pd_\nu \phi) - V \lp \phi \rp,
\end{equation}
where $g_{\mu \nu}$ is the metric. 

The focus of this paper is on spherically-symmetric collapse. Following 
\cite{PhysRevD.36.3575,Garfinkle:1994jb,Brady:1997fj}, 
we use a very efficient integration scheme obtained by introducing the retarded time coordinate $u$ and 
radial coordinate $r$, with metric
\begin{equation} \label{eq:metric}
\dd s^2 = - g(u,r) \s \bar{g}(u,r) \s \dd u^2 - 2 \s g(u,r) \s \dd u \s \dd r + r^2 \s \dd \Omega^2.
\end{equation}
As usual, $\dd \Omega^2$ is the metric on the unit sphere, and we suppose that
$g, \bar{g}$ are two smooth functions. 
Without loss of generality, up to a redefinition of $u$, we can impose boundary
conditions at the origin, $\bar{g}(u,0) = 1$. Imposing that there is no conical singularity 
at $r=0$ then implies that $g(u,0) = 1$~\cite{PhysRevD.36.3575}.

We follow the conventions of \cite{PhysRevD.36.3575,Garfinkle:1994jb} and introduce the notation 
$\bar h$ for the scalar field $\phi$. Radial derivatives of $\bar h$ are used to define an auxiliary field
$h$. One can show that the Einstein equations are fully equivalent to a system of first order equations:
\begin{align}
& \pd_u h - \frac{\bar{g}}{2} \s \pd_r h = 
	\frac{h - \bar h}{2 \s r} \s \left[ \lp 1 - 8 \s \pi \s G \s r^2 \s V \lp \bar h \rp \rp g - \bar{g} \right]
	- \frac{g}{2} \s r \s V' \lp \bar h \rp, \label{eq:FE1}\\
& \pd_r \ln (g) = \frac{4 \s \pi \s G}{r} \s (h - \bar h)^2, \label{eq:FE2}\\
& \pd_r \lp r \s \bar{g} \rp
	= \lp 1 - 8 \s \pi \s G \s r^2 \s V \lp \bar h \rp \rp g,\\
& \pd_r \lp r \s \bar{h} \rp=h \label{eq:FE4},	 
\end{align}
The first of these equations is a form of the Klein-Gordon equation which can be integrated using 
the method of characteristics.
Starting from the initial data surface $u=0$, we label the ingoing radial null geodesics
by a coordinate $v$. The ingoing null geodesics for the metric Eq. (\ref{eq:FE1}) satisfy
the characteristic equation for (\ref{eq:FE1}),
\begin{equation}
\left.\partial_u r\right|_v=-{\bar g\over 2}.\label{eq:FEN2}
\end{equation}
Changing to null coordinates, so that $h(u,r)$ becomes $h(u,v)$, gives the evolution
along the characteristic surfaces of constant $v$,
\begin{equation}
 \pd_u h = 
	\frac{h - \bar h}{2 \s r} \s \left[ \lp 1 - 8 \s \pi \s G \s r^2 \s V \lp \bar h \rp \rp g - \bar{g} \right]
	- \frac{g}{2} \s r \s V' \lp \bar{h} \rp, \label{eq:FEN1}
\end{equation}
In order to solve these field equations, we have adapted the numerical procedure 
from Refs.~\cite{PhysRevD.36.3575, PhysRevD.49.890, Garfinkle:1994jb}.   
Starting from given initial data for $\bar h$ and $r$ at $u = 0$, we first compute $h(0,v)$, $g(0,v)$, and $\bar{g}(0,v)$ 
using (\ref{eq:FE2}-\ref{eq:FE4}). We evolve $r$ and $h$ in the $u$ direction using Equations~\eqref{eq:FEN2} and 
\eqref{eq:FEN1}, discarding points for which $r$ becomes negative. At each step, the constraints 
(\ref{eq:FE2}-\ref{eq:FE4}) are solved by integrating in the $v$ direction.

Evolution methods based on the 3+1 space and time coordinates solve their constraints
on the initial time hypersurface, usually as a boundary value problem, and can be subject to 
constraint violation at later times.
This is not an issue with the null coordinate formalism. 
The method only requires us to solve the \emph{ordinary} differential equations, 
~\eqref{eq:FEN1}  and~\eqref{eq:FEN2}, with integrations over $v$ at each time step. 
As a result, the method is remarkably accurate, fast and reliable.

When a black hole forms, it is possible to follow the evolution up to the null surface $u=u_T$ 
which contains a marginally trapped surface at $r = r_T$. We define the final black hole mass $M_H$ as the  
Bondi mass~\cite{PhysRev.128.2851, PhysRevD.55.1977},
\begin{equation} \label{eq:BondiMass}
M_{\! H} = \lim_{u\to u_T}\lim_{v \to \infty} \frac{r}{2 \s G} \s \lp 1 - \frac{\bar{g}}{g} \rp,
\end{equation}
since this is appropriate for null coordinate systems. A Schwarzschild black hole metric with
mass $M$, for example, has $g = 1$, $\bar g=1-2GM/r$ and $M_H=M$.
At the marginally trapped surface $g\to\infty$, and the computational grid has to be compressed to 
counter the growth in the right hand side of \eqref{eq:FEN1}. In practice, the integration is stopped 
when $\bar{g} / g$ reaches a predetermined value. The Bondi mass is calculated at the final
value of $u$ and with $v$ at the extreme edge of the coordinate grid.

Removing the limits from (\ref{eq:BondiMass}) gives a local quantity $M_B(u,v)$ which evolves according to 
\begin{equation}\label{eq:dmdu}
\partial_u M_{\! B} = - 2 \s \pi \s r^2 \s \lp \frac{2}{g} \s \lp \partial_u\bar h \rp^2 + \bar{g} \s V (\bar h ) \rp.
\end{equation}
When $g$, $\bar{g}$, and $V$ are positive, then $\partial_u M_{\! B}  \leq 0$.
We will use $-\partial_u M_{\! B}$ as a measure of the energy flux from the collapsing star.
Any increase of $M$ along in the ingoing null direction indicates (at least if $V$ remains positive) an 
artefact from the numerical integration, and the corresponding runs are discarded.    We can also use (\ref{eq:dmdu}) to 
put bounds on the error in the black hole mass from truncating the integration before the trapped surface at $u=u_T$.
This gives better control of the black hole mass than we would have using the mass at the trapped
surface, $r_T/2G$, which was used in previous work.

\section{Numerical results}

We preface the full analysis with some results on the collapse of a massive, real, scalar field 
without self-interaction. Depending on initial conditions, the system can collapse to a black hole 
or a stable oscillaton, i.e. an oscillating field configuration that maintains its radial profile 
\cite{Seidel:1993zk,Kolb:1993zz}. The phase diagram for relativistic collapse
in terms of mass and radius was obtained semi-analytically 
in Ref.~\cite{Goncalves:1997qp}. 
The fully relativistic collapse of a massive scalar field was studied in 
some detail using null coordinates in Ref.~\cite{Brady:1997fj} and using a 3+1 approach in
Ref. \cite{Alcubierre:2003sx}.  

We use the null coordinate approach to plot the phase diagram in terms of the axion mass $m$,
the initial radius $R$ and Bondi mass $M_{\! B}$.
The choice of initial density profile is somewhat arbitrary, but we choose to work with a Gaussian 
profile which has been used previously for Bose stars \cite{Chavanis:2011jx}. The scalar field
is oscillatory in time, and when projected on to the light-cone in flat space,
\begin{equation} \label{eq:ID1}
\bar h_i(r) = \sqrt{\frac{2 \s M}{\pi^{3/2} \s m^2 \s R^3}} \s \e^{-r^2 / (2 \s R^2)} \cos (m \s r).
\end{equation}
The pre-factor has been chosen so that the mass of the star is $M$ in the non-relativistic limit $Rm\gg1$.
The relationship between the radius and the ingoing null coordinate on the initial surface can be 
specified freely, but the uniform choice $r=2v$ will be used for simplicity. Initial conditions on the
remaining fields are determined by the constraints (\ref{eq:FE2}-\ref{eq:FE4}), which ensure that we 
have a consistent set of initial conditions for the fully relativistic collapse. 

\begin{figure}
\includegraphics[width=0.49\linewidth]{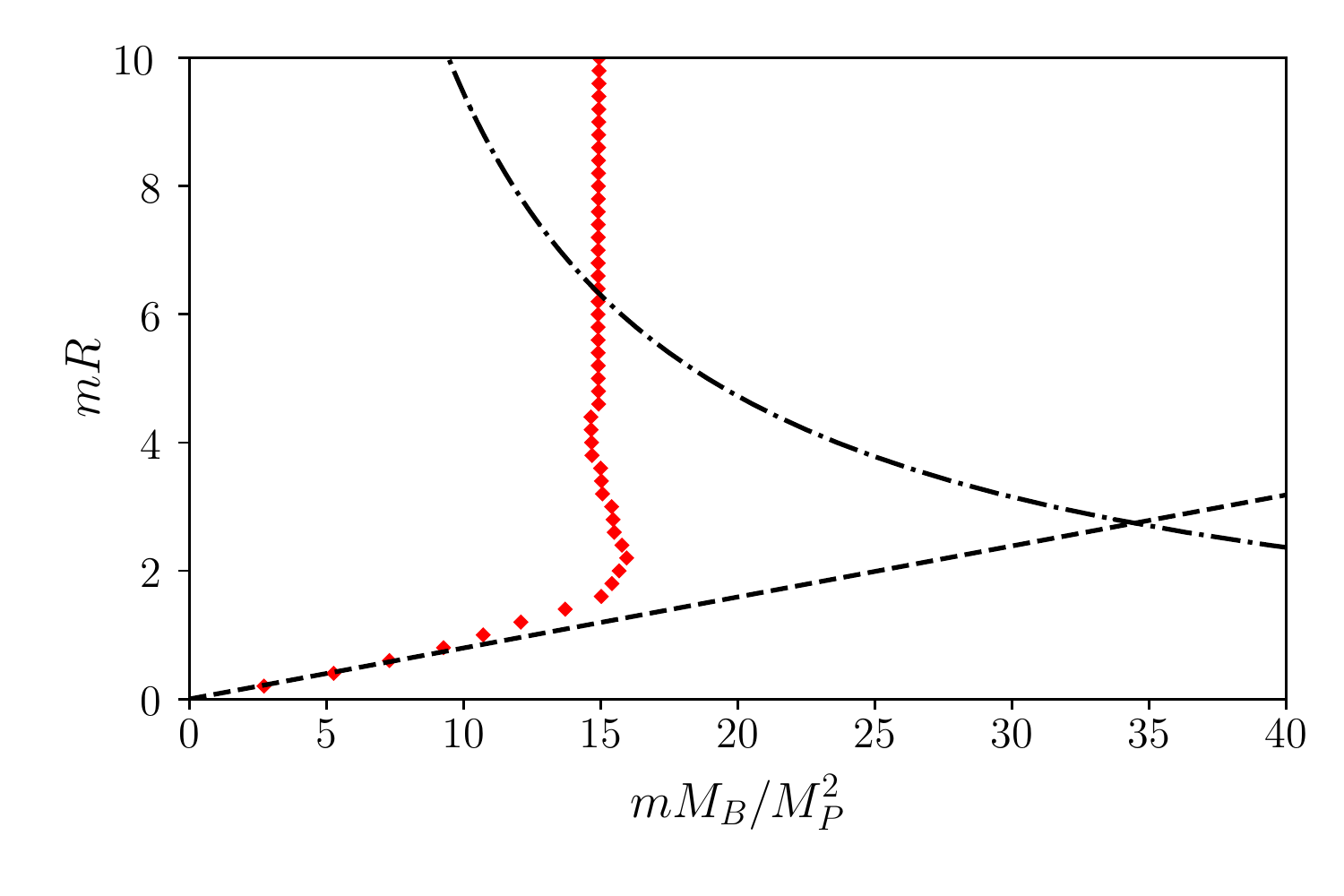}
\caption{
This plot shows the phase plane for the collapse of a massive scalar field without self-interaction,
for an initial profile of the form~\eqref{eq:ID1}. 
The two parameters used are the initial Bondi mass $M_{\! B}$ and the initial radius $R$,
scaled with the axion mass $m$. The dots show the boundary between black hole formation (left) 
and boson stars (right). 
The dashed line shows $R=2GM_{\! B}$, i.e., the mass of a static black hole of radius $R$.
The dash-dot line shows the mass-radius relation for a non-relativistic boson star. 
} \label{fig:findBH_NI}
\end{figure}
The metric and scalar fields are evolved using the method described above.
After dimensional rescaling, the solutions only depend on the initial parameters in the
combinations $mR$ and $mM_B/M_p^2$. 
Figure~\ref{fig:findBH_NI} shows the phase diagram for collapse in terms of these rescaled
parameters. When the mass and radius are appropriate for a stable axion star, the scalar 
field profile settles down to the axion star
field profile within a few oscillations. 
The phase transition boundary is traced out by dots, obtained by a bisection search technique. 
The condition for black hole collapse becomes nearly independent of radius when $m \s R \gtrsim 2$, with critical 
Bondi mass $M_B\approx 15.22 M_p^2m^{-1}$ and mass parameter $M\approx 15.55 M_p^2m^{-1}$. 
The Bondi mass is in very good agreement with the results of the 3+1 approach, where the critical ADM mass is 
$M_{\rm ADM}\approx 15.2 M_p^2m^{-1}$ (after accounting for the switch from Planck mass to 
reduced Planck mass) \cite{Alcubierre:2003sx}.

Inclusion of self-interaction leads to a third possible outcome of gravitational collapse,
a Bosenova, where the collapsing field loses mass in 
pulses of axion radiation. The possibility of collapse due to self-interaction was first noticed in
the non-relativistic limit \cite{Chavanis:2011jx,Chavanis:2016dab}, but the radiation 
is highly relativistic, as pointed out in Ref. \cite{Levkov:2016rkk}. A fully relativistic
treatment of axion collapse with general relativity was given in Ref. ~\cite{Helfer:2016ljl}. The phase plane of mass 
and axion scale was also discussed in~\cite{Helfer:2016ljl}, 
where evidence was given for the existence of a tri-critical point between the three outcomes. 
We make use of our rapid integration scheme to give a clearer picture of the phase plane.

The Gaussian profile is used for the initial data as before, but with fixed initial radius. 
It is desirable to have stable axion stars settle down as quickly as possible to their
final form in order to keep down the integration time. In order to achieve this, we choose the initial radius
using the radius for non-relativistic Bose stars \cite{Chavanis:2011jx,Chavanis:2016dab},
$R\approx 2(24\pi^3)^{1/2}M_p^2/Mm^2$.
Figure \ref{fig:findBH_NI} shows that this radius lies in the region of the phase diagram
where the dependence on radius is weak.

\begin{figure}
\includegraphics[width=0.45\linewidth]{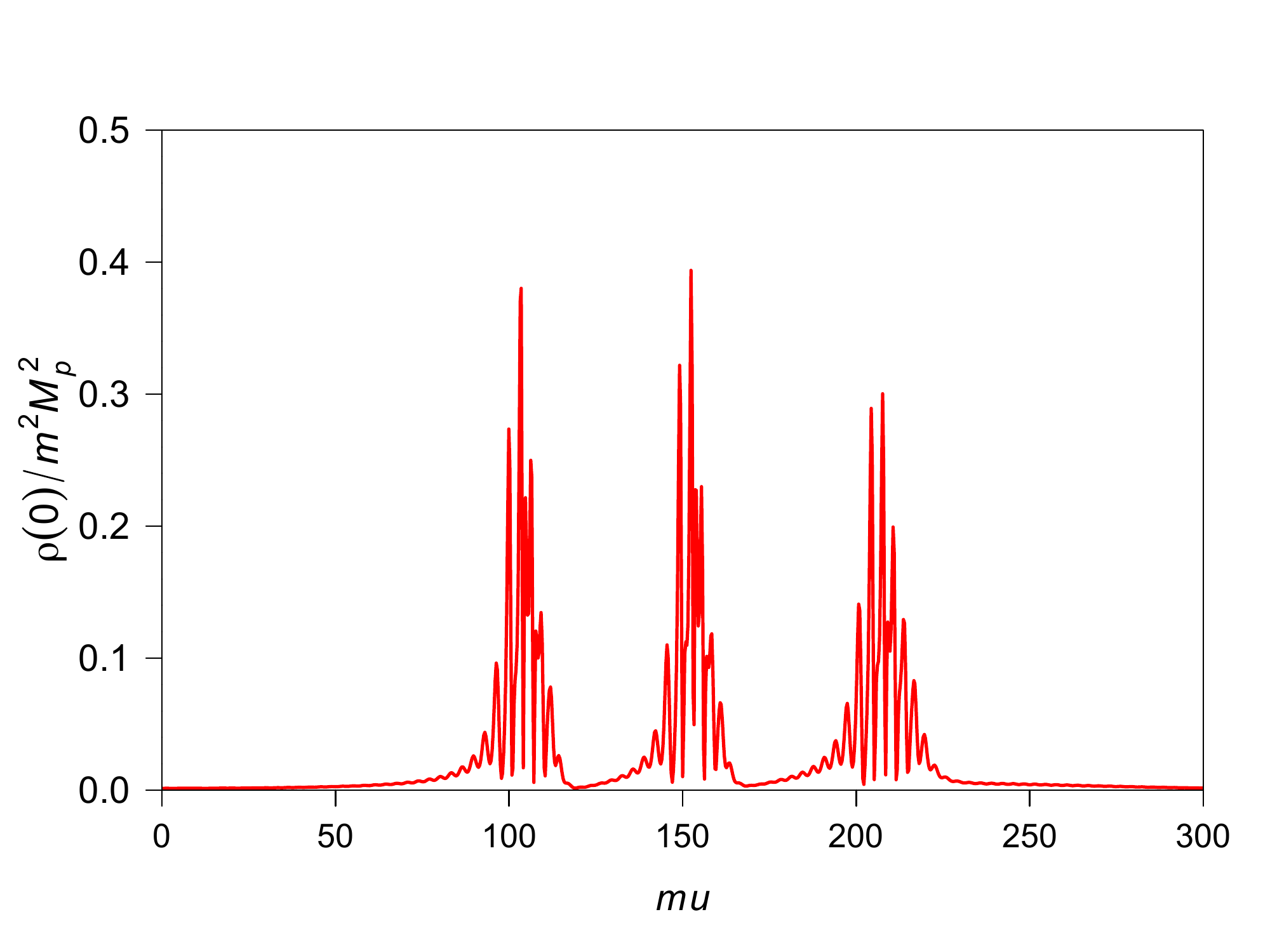}
\includegraphics[width=0.45\linewidth]{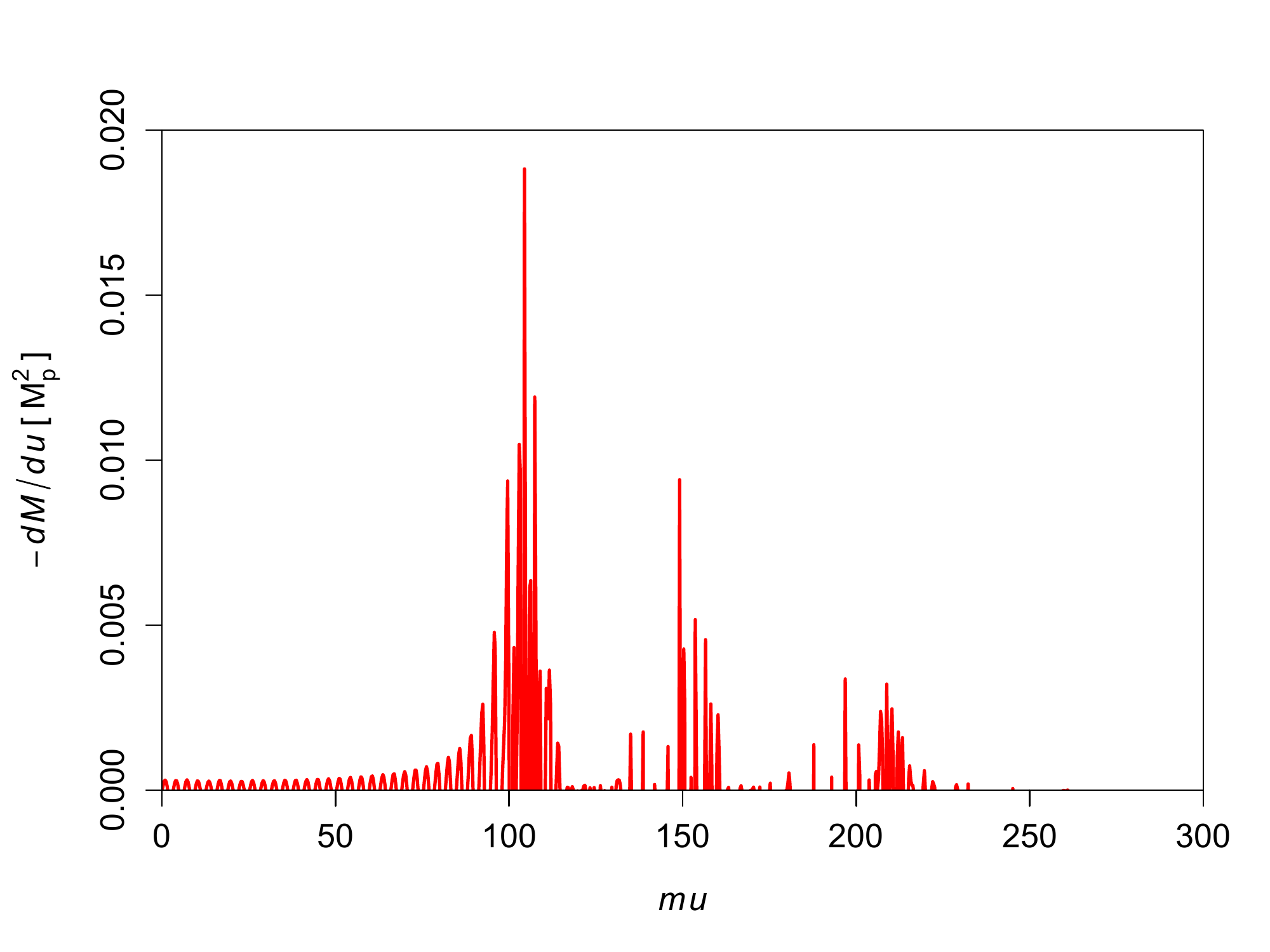}
\caption{The central density (left) and the energy flux (right) for a collapsing
Bosenova are plotted against the retarded time $u$. The flux is measured on 
the outer null edge of the integration volume, $mv=10^3$. In this example
$mM_B/M_p^2=8.9$, where $m$ is the axion mass, $M_B$ is the Bondi mass,
and the axion scale $f=0.1M_p$.  The plot shows density spikes inside the star 
which produce pulses of axion radiation travelling at close to the speed of light.
} \label{fig:flux}
\end{figure}
The Bosenova is characterised by collapses of the stellar core followed by bursts
of axion radiation. The collapse and burst pattern is repeated until a significant portion
of the initial mass has been radiated away. An example is illustrated in figure \ref{fig:flux}, which shows
the central density and the radiation escaping at the edge of the integration volume 
as functions of retarded time. The radiation escapes a short retarded time after each collapse, 
indicating that the radiation is highly relativistic. Burst can be highly irregular, both in their timing
and amplitude. When a black hole forms instead of a Bosenova, this tends to happen 
at the same retarded time as the first spike in the central density.

\begin{figure}
\includegraphics[width=0.5\linewidth]{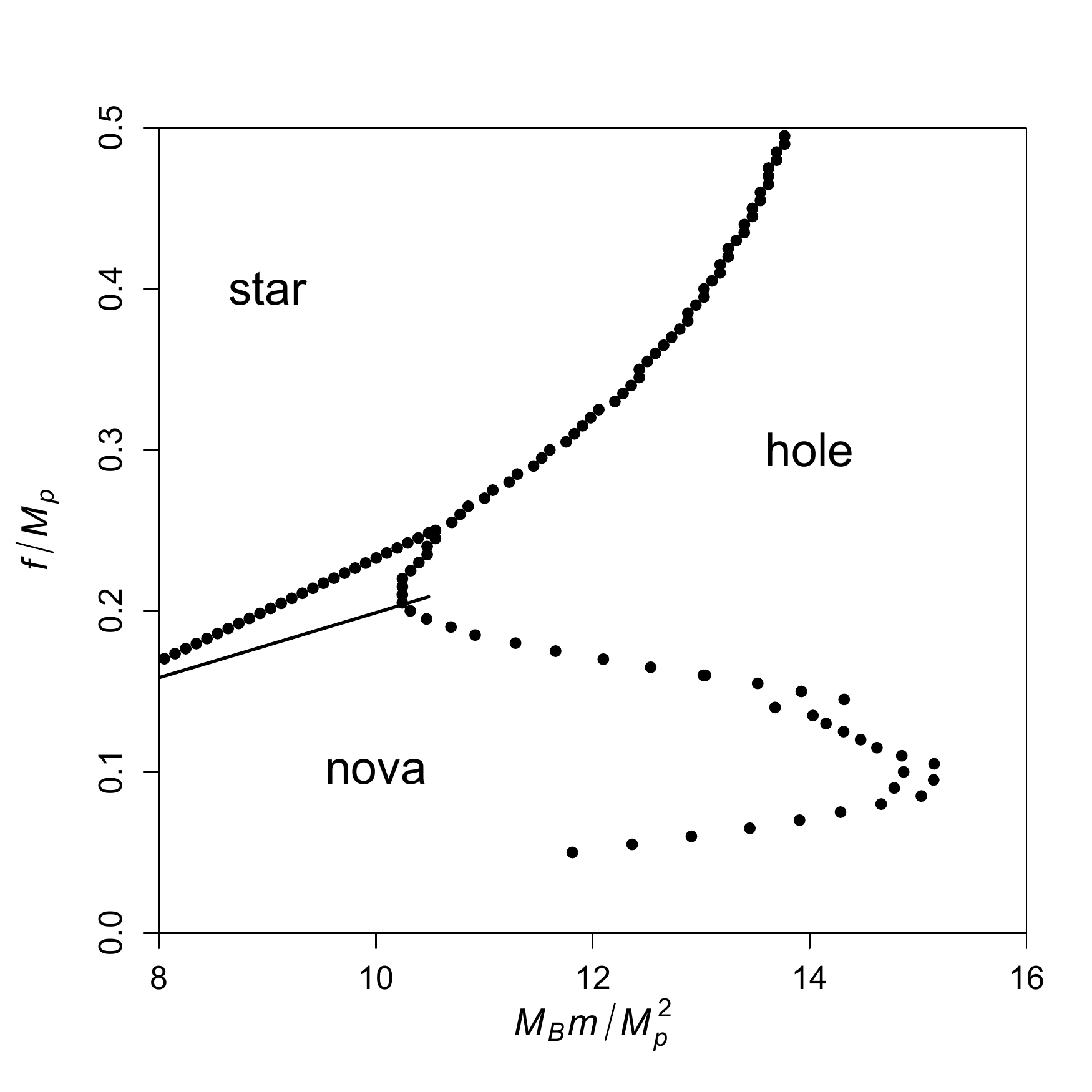}
\caption{Phase diagram for axion scalar field collapse with axion mass $m$, scale parameter $f$ and Bondi
mass $M_B$, with $M_p=(8\pi G)^{-1/2}$. The thin line shows the maximum mass of a
non-relativistic Boson star with quartic self interaction. The boundary between black holes and Bosenovas is diffuse,
and the plot shows only the largest mass initial condition which fails to form a black hole. 
The diagram is compiled using a bisection search technique, with retarded time range 
$u=10^3m^{-1}$. Trapped surface detection uses $\bar g/g=10^{-3}$ and the Bosenova
is defined as a collapse of the core with central density $\rho>10^{-1}m^2M_p^2$. Some points have 
been checked using $u=2\times10^3m^{-1}$ and $\bar g/g=10^{-4}$.
} \label{fig:phase}
\end{figure}
The phase diagram is shown in figure \ref{fig:phase}. There is a tricritical point
in agreement with Ref. \cite{Helfer:2016ljl}, but the mass and axion scale at the tricritical point are
substantially different from the eralier results. The difference is too large to be explained by the
difference between using initial conditions on a null surface instead of
a timelike surface. Our results are broadly consistent with the non-relativistic limit though, where
there is a critical mass for the collapse of a Bose star \cite{Chavanis:2011jx,Chavanis:2016dab},
$M\approx 50.77 \,f M_p /m$, shown on figure \ref{fig:phase}.

\begin{figure}
\includegraphics[width=0.45\linewidth]{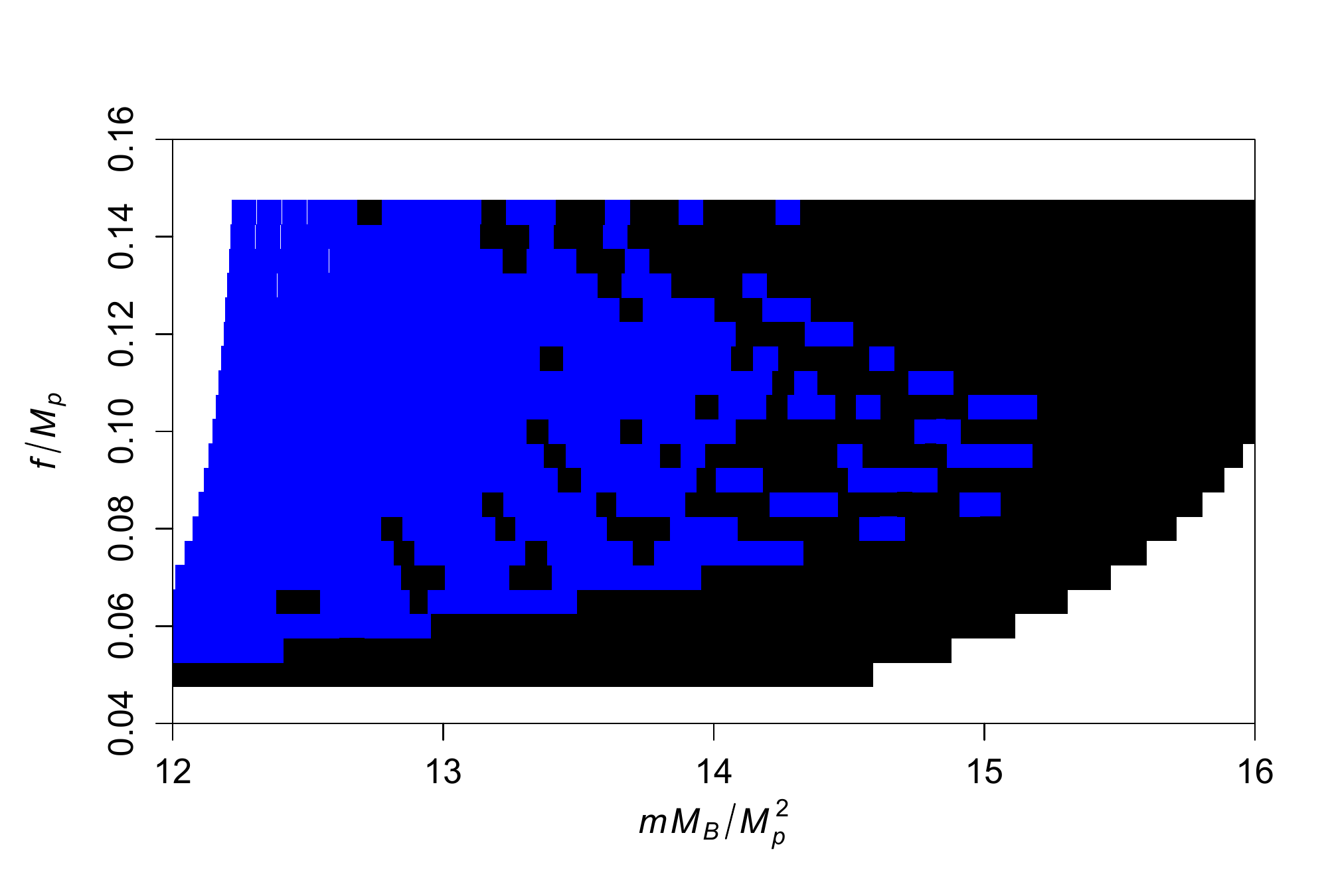}
\caption{Part of the phase diagram where the phase boundary is diffuse.  Each black pixel denotes
a set of initial conditions which forms a black hole, and each blue pixel denotes a set of initial conditions
which forms a Bosenova.   
} \label{fig:batch}
\end{figure}

The phase boundary between axion stars and black holes is sharply
defined, and the mass of the black hole is discontinuous at the phase boundary.
However, the phase boundary between black holes and Bosenovas becomes diffuse
at small values of the axion scale parameter $f$, in the sense that there is a range of 
masses near the phase boundary where the outcome of gravitational collapse can go either way, 
as shown in figure \ref{fig:batch}. Some of the axion clumps with initial conditions near 
the phase boundary emit enough axion radiation to avoid forming a black hole.
This seems to happen erratically. A similar phenomenon was observed for 
the collapse of non-self interacting scalar fields in Ref.  \cite{Alcubierre:2003sx}, 
where the effect was ascribed to gravitational cooling, using a term introduced for scalar field
emission by collapsing Boson stars in Ref. \cite{Seidel:1993zk}. 

\section{Discussion}

Gravitational collapse with nothing more than gravity and a scalar field is
a remarkably rich subject. It seems sensible to build up an understanding of it
in small steps, the simplest being spherically symmetric collapse.
We have considered three possible scenarios for axion collapse: axion stars, black holes and 
Bosenovas. The numerical results clearly point to a critical point with Bondi mass 
$M_B\approx 10.6 M_p^2m^{-1}$ and axion decay constant $f\approx 0.25 M_p$
when the initial conditions are presented on a phase diagram.
Mostly, the distinction between the different phases is clear, but in some parts of the 
phase diagram, there is no clean line between initial conditions which collapse to a black hole 
and those which remain non-singular. There may also exist special
final states like the self-similar solutions for massless scalar collapse \cite{Choptuik:1992jv}
which we have not considered.

The fate of a Bosenova is to eject mass until it eventually settles down into a
stable axion star. In terms of the eventual outcome, the stars and Bosenova's
are similar. However, the difference has a large physical significance for
dark energy scenarios, since the axion radiation from many bursts and many
sources would combine into a background of incoherent ALP particles. 

Just how dependent the results are on the initial density profile and the use of null initial
data surface may be determined through further work. The Gaussian initial density profile is
known to work to within a few percent for results on non-relativistic Bose stars \cite{Chavanis:2016dab},
but we have done some runs with radically different density profiles and find $O(1)$ changes in the
masses at the phase boundaries.

The null coordinate approach is fast, accurate and reliable, but it is limited to spherically symmetric
collapse. It provides an important check on the results obtained from
more sophisticated 3+1 integration schemes, but ultimately the 3+1 schemes are necessary for
handling general collapse situations. 
We hope the clearer view of the simplest-case scenario considered here will help guiding future studies in this direction. 

\acknowledgments
We would like to acknowledge useful discussions with Ruth Gregory and Gerasimos Rigopoulos.
We are grateful for the hospitality of the Perimeter Institute, Waterloo, Canada where
this paper was written. IGM and FM are supported by the Leverhulme grant RPG-2016-233. IGM also 
acknowledges some support from the Science and Facilities Council of the 
United Kingdom grant number ST/P000371/1.

\bibliography{biblio}

\end{document}